\documentclass[12pt]{iopart}
\pdfoutput=1 
\usepackage{iopams}
\usepackage[usenames,dvipsnames]{color}
\usepackage{graphicx}

\begin{document}
\title{
On superstatistics of energy for a free quantum Brownian particle
}
\author{J. Spiechowicz and J. {\L}uczka}
\address{Institute of Physics and Silesian Center for Education and Interdisciplinary Research, University of Silesia, 41-500 Chorz{\'o}w, Poland}
\ead{j.spiechowicz@gmail.com}
\begin{abstract}
We consider energetics of a free quantum Brownian particle coupled to thermostat of temperature $T$ and study  this problem in terms of the lately formulated  quantum analogue of the energy equipartition theorem. We show how this quantum counterpart can be derived from the Callen-Welton fluctuation-dissipation relation and rephrased in terms of superstatistics.  We analyse the influence of the system-thermostat coupling strength and the memory time of the dissipation kernel on statistical characteristics of the particle energy and its specific heat. 

\end{abstract}

\maketitle

\section{Introduction}
Is any simpler example of a dissipative quantum system than a free Brownian particle? 
No and it could seem that (almost) all is known on this problem and nothing new can be presented for this hackneyed setup.  We want to demonstrate that there still remain new results to be ferret out for this system. It is a consensus that any quantum system coupled to thermostat reaches stationary state which is thermodynamic equilibrium one. However, an  explicit form of this state is not known in a general case. Some partial results have been obtained for some selected models, in some limiting cases and some asymptotic regimes. We intentionally overused the word "some" to emphasize that our knowledge on thermodynamic states of quantum systems is still very limited. The intention of this paper is to present various new aspects of averaged energy of a free quantum Brownian particle. 

In Sec. 2 we shortly review the main results on modelling of the quantum dissipative particle in terms of the Generalized Quantum  Langevin Equation (GQLE) and recall  the relation for the quantum 
analogue of energy equipartition theorem. In Sec. 3 we show how this relation can be derived from the fluctuation-dissipation theorem of the Callen-Welton type. In Sec. 4, we reformulate this relation in the framework of superstatistics, both in the frequency and energy domain. In Sec. 5, we analyse the corresponding probability distribution in the domain of thermostat oscillators energy.
Sec. 7 is devoted to statistical moments of the energy of the Brownian particle and in Sec. 8 the specific heat is analyzed. The paper ends with with a summary in Sec. 8. In two Appendices we present the solution of GQLE and examples of the dissipation (memory) function appearing in GQLE which are considered in this paper.

\section{Model of dissipation for quantum Brownian particle} 
The free quantum Brownian particle $S$ of mass $M$ coupled to thermostat is described by the Caldeira-Leggett Hamiltonian   \cite{maga,uler,caldeira,ford,gaussian,et2,weis}: 
\begin{equation} \label{H}
H=\frac{p^2}{2M} + \sum_i \left[ \frac{p_i^2}{2m_i} + \frac{m_i
\omega_i^2}{2} \left( q_i - \frac{c_i}{m_i \omega_i^2} x\right)^2 \right]  
\end{equation}
in which thermostat $E$ is modelled as a set of harmonic oscillators in an  thermodynamic equilibrium  state of temperature $T$.  
The operators  $\{x, p\}$ are the coordinate and momentum operators of the Brownian particle and $\{q_i, p_i\}$ refer to  the coordinate and momentum operators of the $i$-th thermostat oscillator of mass $m_i$ and the eigenfrequency $\omega_i$. The parameter $c_i$ characterizes the interaction strength of the particle with the $i$-th oscillator. There is the counter-term, the last term proportional to $x^2$, which is included to cancel a harmonic contribution to the particle potential. All coordinate and momentum operators obey canonical equal-time commutation relations. 
 
From  the Heisenberg equations of motion for all coordinate and momentum operators $\{x, p, q_i,p_i\}$ one can  obtain an effective equation of motion only for the particle coordinate $x(t)$ and momentum $p(t)$. It is called a Generalized Quantum Langevin Equation and for the momentum operator of the Brownian particle it reads \cite{bialasPRA}
\begin{equation}\label{GLE2}
{\dot p}(t)
+\frac{1}{M} \int_0^t \gamma(t-s) p(s) \, ds = -\gamma(t) x(0)+ \eta(t),  
\end{equation}
where  dot denotes time derivative and 
 $\gamma(t)$ is the dissipation function (damping or memory kernel), 
\begin{equation} \label{gamma}
\gamma(t) =\sum_i \frac{c_i^2}{m_i \omega_i^2} \cos(\omega_i t)  \equiv 
\int_0^{\infty}  J(\omega) \cos(\omega t)  d \omega  
\end{equation}
and 
\begin{eqnarray} \label{spectral}
J(\omega) = \sum_i \frac{ c_i^2}{ m_i \omega_i^2} \delta(\omega -\omega_i) 
\end{eqnarray} 
is a spectral function of the thermostat which contains information on its modes and the Brownian particle-thermostat interaction. The term $\eta(t)$ can be interpreted as a random force acting on the Brownian particle, 
\begin{equation} 
\eta(t) =\sum_i c_i \left[q_i(0) \cos(\omega_i t) + \frac{p_i(0)}{m_i \omega_i}\sin(\omega_i t) \right], 
\label{force} 
\end{equation}
which  depends  on initial conditions $\{q_i(0), p_i(0)\}$ imposed on the thermostat oscillators.   
We assume the factorized initial state  of the composite  system $S+E$, i.e., $\rho(0)=\rho_S(0)\otimes\rho_E(0)$, where $\rho_S(0)$ is an arbitrary state of the Brownian particle and $\rho_E(0)$ is an equilibrium canonical state of the thermostat 
of temperature $T$, namely,
\begin{eqnarray}
\rho_E(0) = \mbox{exp}(-H_E/k_B T)/\mbox{Tr}[\mbox{exp}(-H_E/k_B T)],
\end{eqnarray}
where:
\begin{equation}
H_E= \sum_i \left[ \frac{p_i^2}{2m_i} + \frac{1}{2} m_i \omega_i^2 q_i^2 \right] 
\end{equation}
is the Hamiltonian of the thermostat (quantum environment). The factorization means that there are no initial correlations between the particle and  the thermostat.
The solution of Eq. (\ref{GLE2})  reads (see Appendix A)
\begin{eqnarray}\label{p(t)} 
p(t)= R(t)p(0) - \int_0^t  R(t-u) \gamma(u) du\; x(0)+ \int_0^t R(t-u) \eta(u) du,  
\end{eqnarray}
where $R(t)$ is a response function determined by its Laplace transform,  
\begin{equation}\label{RL} 
\hat{R}_L(z) = \frac{M}{Mz + \hat \gamma_L(z)}. 
\end{equation}  
Here, $\hat \gamma_L(z)$ is a Laplace transform of the dissipation function $\gamma(t)$ and for any function  $f(t)$ its Laplace transform is defined as 
\begin{equation}\label{fL} 
\hat f_L(z) = \int_0^{\infty} {\mbox e}^{-zt} f(t) dt. 
\end{equation}  

Using Eq. (\ref{p(t)}), one can  calculate averaged kinetic energy 
$\langle E(t) \rangle =\langle p^2(t)\rangle/2M$ of the Brownian particle. It is of course a total average energy of the particle. 
In the thermodynamic limit for the thermostat (which is  infinitely extended) and in the long time limit $t \to \infty$, when a thermal equilibrium state is reached, the mean kinetic energy $\langle E \rangle$ can be presented in the form (for detailed derivation, see Ref. \cite{bialasPRA})
\begin{equation}\label{Ek}
 \langle E \rangle = \int_0^{\infty} \mathcal{E}(\omega)\mathbb{P}(\omega)  d\omega ,  
\end{equation}
where  
\begin{equation}\label{ho}
\mathcal{E}(\omega) = \frac{\hbar \omega}{4} 
\coth\left({\frac{\hbar \omega}{ 2k_BT}}\right)
\end{equation} 
and 
\begin{eqnarray}\label{P}
\mathbb{P}(\omega) = \frac{1}{\pi} \left[\hat{R}_L(i\omega) + \hat{R}_L(-i\omega) \right]. 
\end{eqnarray}
The function $\mathbb{P}(\omega)$ fulfils all  conditions imposed on the {\it probability density}: (i) it is non-negative, i.e.  $\mathbb P(\omega)\ge 0$, and  (ii) normalized on the frequency half-line, i.e.  $\int_0^{\infty} d\omega \; {\mathbb P}(\omega) = 1$. The proof is presented in Ref. \cite{bialasJPA}

$\mathcal{E}(\omega)$ is an equilibrium kinetic energy per one degree of freedom of the thermostat of temperature $T$ at initial time $t = 0$ \cite{feynman}, namely 
\begin{eqnarray} \label{rho}
\mathcal{E}(\omega) = \mbox{Tr} \left[  \frac{p_i^2}{2m_i} \; \rho_E(0)\right].   
\end{eqnarray} 
However, we note that it is not equal to the mean kinetic energy of a single thermostat oscillator coupled to the Brownian particle in the long time regime, i.e. when the total system $S+E$ approaches the thermodynamic equilibrium state.
The r.h.s. of Eq. (\ref{Ek}) is then an averaging over frequencies $\omega$ of those thermostat oscillators which contribute to $\langle E \rangle$ according to the probability distribution (\ref{P}). One can attempt to interpret the relation (\ref{Ek}) in the following way: the Brownian particle "remembers" the initial state of the thermostat even in the long time regime because its averaged kinetic energy $\langle E \rangle$ for $t\to\infty$ is the mean kinetic energy $\mathcal{E}(\omega)$ of the thermostat oscillators at initial time $t=0$. In this sense the Brownian particle has an infinitely long memory with respect to the energy.

For comparison with results derived in the next section, we rewrite the formula (\ref{P}) in another form.  The Laplace transform $\hat{\gamma}_L(z)$ of the memory function $\gamma(t)$ in Eq. (\ref{RL}) can be expressed by the cosine and sine Fourier transforms,  
\begin{eqnarray}
\hat{\gamma}_L(i\omega) &= \int_0^{\infty}  \gamma(t)  \mbox{e}^{-i\omega t} dt = A(\omega) - i B(\omega) \label{L-F}\\
A(\omega) &= \int_0^{\infty}  \gamma(t) \cos{(\omega t)} dt, \label{cos}\\  
B(\omega) &= \int_0^{\infty}  \gamma(t) \sin{(\omega t)} dt. \label{sin}
\end{eqnarray}
Next, we insert them into Eqs. (\ref{RL}) and (\ref{P}) and obtain an equivalent form of the probability distribution, namely, 
\begin{equation} \label{Pp}
\mathbb{P}(\omega) = \frac{2 M}{\pi} \frac{ A(\omega)}{A^2(\omega)+[B(\omega)-M\omega]^2}.
\end{equation}
Note that, via Eq. (\ref{gamma}),  the function $A(\omega)$ is related to the spectral function $J(\omega)$. Because the latter is non-negative, $J(\omega) \ge 0$, and the denominator in (\ref{Pp}) is positive, the function $\mathbb{P}(\omega)$ is non-negative as required for a probability density. 

\section{Derivation from fluctuation-dissipation theorem}

If we know the representation (\ref{Ek}) then $\mathbb{P}(\omega)$ can formally be obtained from the fluctuation-dissipation theorem of the Callen-Welton type \cite{weis,call,zubarev,landau}. We recall this relation for a special case of the  momentum operator. It takes the form (see Eq. (124.10) in Ref. \cite{landau}),  
\begin{equation}
	\label{landau}
	\langle p^2 \rangle = \frac{\hbar}{\pi}\int_0^\infty d\omega\, \coth{\left[\frac{\hbar\omega}{2k_BT}\right]} \,\tilde\chi''(\omega),
\end{equation}
where $\tilde\chi''(\omega)$ is the imaginary part of the generalized susceptibility $\tilde\chi(\omega) = \tilde\chi'(\omega) + i\tilde\chi''(\omega)$.  In turn, the susceptibility $\tilde\chi(\omega)$ is a Fourier transform 
\begin{equation}
	\label{four}
	\tilde\chi(\omega) = \int_{-\infty}^{\infty} \mbox{e}^{i\omega t} \chi(t) dt
\end{equation}
of the retarded thermodynamic Green function \cite{zubarev} (see also Eq. (126.8) in the Landau-Lifshitz book \cite{landau}),  
\begin{equation}
	\label{green} 
	\chi(t) =  \frac{i}{\hbar} \theta(t) \langle [p(t), p(0)] \rangle,
\end{equation}
where $\theta(t)$ is the Heaviside step function,  $p(t) = \exp(i{ H} t/\hbar) p(0) \exp(-i{ H} t/\hbar)$ and  averaging  is over the Gibbs canonical statistical operator $\rho \propto \mbox{exp}[- H/k_B T]$ with $H$ given by Eq. (\ref{H}).  

{\it Remark}: In the linear response theory, $\chi(t)$  is also called a response function which,  however,   is not the same as the response function $R(t)$ in Eq. (\ref{p(t)}). To avoid confusion, we accept  an equivalent name as the retarded thermodynamic Green function \cite{zubarev}.   
\\\noindent
Using Eqs.  (\ref{force}) and (\ref{p(t)}) one  can calculate the commutator in Eq. (\ref{green}) and the Green function (\ref{green}) reads
\begin{equation} 	\label{comut} 
			\chi(t) =   \theta(t) \int_0^t R(t-u) \gamma(u) du. 
\end{equation}
It is a convolution of two functions $R(t)$ and $\gamma(t)$ and its Fourier transform 
is a product of the Fourier transforms of $R(t)$ and $\gamma(t)$. Because of the $\theta$-function, the Fourier transform of $\chi(t)$ in Eq. (\ref{comut}) can be expressed by Laplace transforms of $R(t)$ and $\gamma(t)$, namely,   
\begin{equation} 	\label{xiF} 
			\tilde\chi(\omega) =   \hat R_L(-i\omega) \hat \gamma_L(-i\omega).  
\end{equation}
Now, we can exploit the representation (\ref{L-F})  to get the generalized susceptibility
\begin{equation} \label{xiAB}
\tilde\chi(\omega) 
 =  M \;  \frac{ A^2(\omega) +  B^2(\omega) -M\omega  B(\omega) + 
iM\omega  A(\omega)}{A^2(\omega)+[B(\omega)-M\omega]^2}. 
\end{equation}
We compare Eq. (\ref{Ek}) with Eq. (\ref{landau}) and make use of the imaginary part $\tilde\chi''(\omega)$ of the susceptibility $\tilde\chi(\omega)$ in the above equation to obtain  the expression  
\begin{equation} \label{chi}
\mathbb{P}(\omega) = \frac{2}{\pi M \omega } \, \tilde\chi''(\omega) 
= \frac{2 M}{\pi} \frac{ A(\omega)}{A^2(\omega)+[B(\omega)-M\omega]^2}, 
\end{equation}
which is the same as Eq. (\ref{Pp}). However, in the framework of the linear response theory we still are not able to prove the  normalization of $\mathbb{P}(\omega)$.

\section{Superstatistics point of view} 

The relation (\ref{Ek}) has been  investigated in Ref. \cite{bialasPRA} for various dissipation mechanisms modelled by the memory kernel $\gamma(t)$. Mainly the probability distribution $\mathbb P(\omega)$ and mean energy $\langle E \rangle$ has been studied there. Here, we attempt to look at this issue from different perspective. The relation (\ref{Ek}) resembles superstatistics: the statistics of the statistics \cite{beck, metzler, magdziarz}. Eq. (\ref{Ek}) is a superposition of two averaging. The first one is over the canonical Gibbs state (\ref{rho}) for the thermostat free (non-interacting with the Brownian particle) oscillators and it yields the averaged kinetic energy of the single thermostat oscillator, i.e. $\mathcal{E}(\omega)$. The second one is over randomly distributed frequencies $\omega$ of the thermostat oscillators according
to the probability density $\mathbb{P}(\omega)$ in which the interaction with a Brownian particle is exactly included. The relation (\ref{Ek}) is formulated in the frequency domain. From the point of view of the probability theory, one can say that there exists a random variable $\xi$ which takes non-negative values, $\xi \in [0, \infty)$,  for which $\mathbb P_{\xi}(\omega) = \mathbb P(\omega)$ 
is its probability density. This random variable is interpreted as a frequency of thermostat oscillators. The relation 
\begin{equation} \label{random}
\zeta = \mathcal{E}(\xi) = \frac{\hbar \xi}{4} 
\coth\left({\frac{\hbar \xi}{ 2k_BT}}\right)
\end{equation}
defines a new random variable $\zeta$ and we can find its probability distribution. Because $ \mathcal{E}(\omega)$ is thermally averaged kinetic energy of thermostat oscillators per one degree of freedom we can recast the problem into the energy representation. For this purpose we rewrite Eq. (\ref{Ek}) in the following way
\begin{eqnarray} \label{pE}
\langle E \rangle &=&  \int_0^{\infty} \mathcal{E}(\omega) \mathbb{P}(\omega) \, d\omega = \int_{\mathcal{E}_0}^{\infty} \mathcal{E} \, \mathbb{P}(\omega(\mathcal{E})) \frac{d\omega}{d\mathcal{E}} \, d\mathcal{E} \nonumber\\ &=& \int_{\mathcal{E}_0}^{\infty} \mathcal{E} \, f_T(\mathcal{E}) \, d\mathcal{E},
\quad \quad \mathcal{E}_0= \frac{k_B T}{2},
\end{eqnarray}
where 
\begin{eqnarray}\label{fE}
f_T(\mathcal{E}) = \mathbb{P}(\omega(\mathcal{E})) \; \frac{d\omega}{d\mathcal{E}}  
\end{eqnarray}
is the probability distribution of the random variable $\zeta$ 
and $\omega =\omega(\mathcal{E})$ is the inverse of the mapping $\mathcal{E} = \mathcal{E}(\omega)$ given by \mbox{Eq. (\ref{ho})}. For the convenience of the reader we depict $\mathcal{E}(\omega)$ in  Fig. 1. It is a one-to-one function which can be uniquely inverted to obtain $\omega(\mathcal{E})$. For $\omega=0$ it assumes the minimal value $\mathcal{E}(0) = \mathcal{E}_0 = k_B T/2$ and for large values of $\omega$ it behaves like a linear function $\mathcal{E}(\omega) \sim (\hbar/4) \omega$. The derivative $d\omega/d\mathcal{E}$ in Eq. (\ref{fE}) can be obtained from the relation  
\begin{equation} \label{dE}
\frac{d\mathcal{E}}{d\omega} = \frac{\hbar}{4}\left\{ \coth\left({\frac{\hbar \omega}{ 2k_BT}}\right) + \frac{\hbar\omega}{2k_BT} \left[1-  \coth^2\left({\frac{\hbar \omega}{ 2k_BT}}\right)\right]\right\}, 
\end{equation}
where $\omega=\omega(\mathcal{E})$ is determined by the inverse of the mapping (\ref{ho}). 
The formula (\ref{pE}) corresponds to the idea of superstatistics in the energy domain: the random variable $\zeta=\mathcal{E}(\xi)$ is the kinetic energy of the thermostat oscillator and its averaging over the probability distribution $f_T(\mathcal{E})$ yields the mean kinetic energy of the Brownian particle $\langle E \rangle$. 
Thermostat oscillators of various kinetic energies contribute to $\langle E \rangle$ in a different degree which is described by the probability density $f_T(\mathcal{E})$. As we will show later on, $f_T(\mathcal{E})$ carries interesting physical content about environment of the studied system.

For any model of dissipation determined by the memory function $\gamma(t)$ the energy probability distribution $f_T(\mathcal{E})$ has the following mathematical properties: (i) $f_T(\mathcal{E})$ is defined on the interval $[\mathcal{E}_0,  \infty)$, where $\mathcal{E}_0$ is given in  Eq. (\ref{pE}), (ii) $f_T(\mathcal{E}) \to \infty$ when $\mathcal{E}\to \mathcal{E}_0$, (iii) $f_T(\mathcal{E}) \to 0$  when $\mathcal{E} \to \infty$, (iv) $\int_{\mathcal{E}_0}^\infty f_T(\mathcal{E})\,d\mathcal{E} = 1$. It is instructive to see that for a free quantum Brownian particle the frequency probability distribution $\mathbb{P}(\omega)$ expressed by Eq. (\ref{P}) depends on the parameters describing the particle (its mass $M$) as well as details of the system-thermostat coupling given by the memory kernel $\gamma(t)$. The latter function is characterized by the coupling strength $\gamma_0$ and the memory time $\tau_c$, see Appendix B.  On the other hand, the energy probability distribution $f_T(\mathcal{E})$ does additionally depend on thermostat temperature $T$. We explicitly denote this fact by putting the letter $T$ into the subscript of the corresponding probability distribution. This observation may be seen as a consequence of performing the intermediate averaging over the thermal equilibrium Gibbs state for thermostat oscillators.
\begin{figure}[t]
	\centering
    \includegraphics[width=0.49\linewidth]{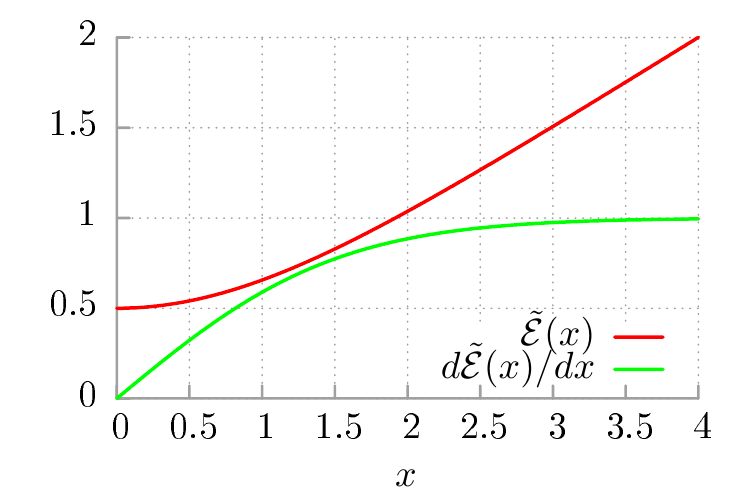}
    \caption{(color online): The rescaled thermal kinetic energy of the thermostat oscillator $\tilde{\mathcal{E}}(x) = \mathcal{E}(x)/k_B T = (1/2) x \coth{x}$ (red curve) together with its first derivative $d \tilde{\mathcal{E}}(x)/dx = \coth{x} + x(1- \coth^2{x})$ (green curve), where $x=\hbar \omega/2 k_B T$.}
    \label{fig1}
\end{figure}

Let us remind that in the classical case the particle momentum (or velocity) is distributed according to the Gaussian probability density and using the same method as in Eq. (\ref{pE}) one can show that the corresponding energy probability density $f_{cl}(E)$ is given by the Gamma distribution
\begin{equation} \label{Ecl}
f_{cl}(E) = \left(\pi k_B T E\right)^{-1/2} \mbox{e}^{-E/k_B T}
\end{equation}
and 
\begin{equation}
	\langle E \rangle = \int_0^\infty E\,f_{cl}(E)\,dE. 
\end{equation}
The density $f_{cl}(E)$ is always a monotonically decreasing function from infinity to zero on its support $[0, \infty)$. Moreover, as it is seen from Eq. (\ref{Ecl}), the classical energy probability distribution $f_{cl} (E)$ does not depend on the particle mass $M$, the coupling constant $\gamma_0$ and the memory time $\tau_c$. It is a drastic difference between the classical and quantum realm. 
We have to stress that $f_{cl}(E)$ is the probability  distribution of energy of the classical Brownian particle while $f_T(\mathcal{E})$ is the probability density of the thermally averaged kinetic energy of thermostat quantum oscillators. 

\section{Analysis of the energy probability distribution}
In the following section we will consider two models of the dissipation mechanism, namely, the Drude and Lorentzian ones. They are characterized by the memory kernel $\gamma(t)$ or equivalently by the spectral density $J(\omega)$ via the cosine Fourier transform, c.f. Eq. (\ref{gamma}). The Drude $\gamma_D(t)$ and Lorentzian $\gamma_L(t)$ models are defined by the dissipation functions 
\begin{equation}
	\label{g-drude}
	\gamma_D(t) = \frac{\gamma_0}{2 \tau_c}e^{-t/\tau_c}, \quad \gamma_L(t) = \frac{\gamma_0}{\pi}\, \frac{\tau_c}{t^2 + \tau_c^2}
\end{equation}
with two parameters $\gamma_0 > 0$ and $\tau_c > 0$. The first, $\gamma_0$, is the particle-thermostat coupling strength and the second, $\tau_c$, is the memory time which characterizes the degree of non-Markovianity in dynamics of the Brownian particle. In this scaling   the functions $\gamma_D(t)$ and $\gamma_L(t)$ tend to the Dirac delta when the memory time goes to zero $\tau_c \to 0$ (the Markovian regime). Then the integral term in the generalized Langevin equation (\ref{GLE2}) reduces to the frictional force of the Stokes form. The corresponding spectral densities take the form 
\begin{equation}
	\label{j-drude}
	J_D(\omega) = \frac{\gamma_0}{\pi}\frac{1}{1 + \omega^2\tau_c^2}, \quad 
	J_L(\omega) = \frac{\gamma_0}{\pi}e^{-\omega \tau_c}.
\end{equation}
From Eq. (\ref{P}) one can derive explicit expressions for the frequency probability density $\mathbb{P}(\omega)$ \cite{bialasPRA}. For the Drude model it reads
\begin{equation} \label{P_D}
	\mathbb{P}_D(\omega) = \frac{1}{\pi} \, \frac{\mu_0\varepsilon^2}{\omega^4 
	+ \varepsilon (\varepsilon -\mu_0) \omega^2 + \mu_0^2\varepsilon^2/4}, 
\end{equation}
For the Lorentzian dissipation one obtains
\begin{equation}
	\label{p-cauchy0}
	\mathbb{P}_L(\omega) = \frac{4\pi}{\mu_0} \, 
	\frac{e^{- \omega/\varepsilon}}{\pi^2 e^{-2\omega/\varepsilon} + c^2(\omega)},
\end{equation}
where 
\begin{eqnarray}
	c(\omega) = e^{-\omega/\varepsilon} \mbox{Ei}(\omega/\varepsilon) - e^{\omega/\varepsilon} \mbox{Ei}(-\omega/\varepsilon) -\frac{2\pi}{\mu_0} \omega 
\end{eqnarray}
and $\mbox{Ei}(z)$ is the exponential integral defined as
\begin{equation}
\mbox{Ei}(z) = -\int_{-z}^\infty \frac{e^{-t}}{t} \, dt.
\end{equation}
There are two control parameters $\varepsilon$ and $\mu_0$ possessing the unit of frequency or equivalently two time scales. The first is the memory time $\tau_c = 1/\varepsilon$. The second is $\tau_v = M/\gamma_0 = 1/\mu_0$ which determines the rescaled coupling strength and in the case of a classical free Brownian particle defines the velocity relaxation time. However, if we recast all quantities into the corresponding dimensionless form then it turns out that both the frequency probability distributions $\mathbb{P}_D(\omega)$ and $\mathbb{P}_L(\omega)$ depend only on their ratio, namely
\begin{equation}
\alpha = \frac{\tau_v}{\tau_c} = \frac{\epsilon}{\mu_0} = \frac{M}{\gamma_0 \tau_c}.
\end{equation}
%
\begin{figure}[t]
	\centering
	\includegraphics[width=0.49\linewidth]{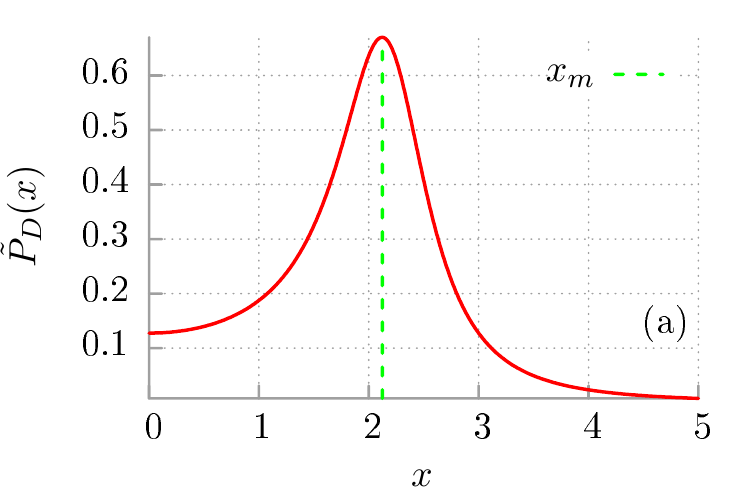}
	\includegraphics[width=0.49\linewidth]{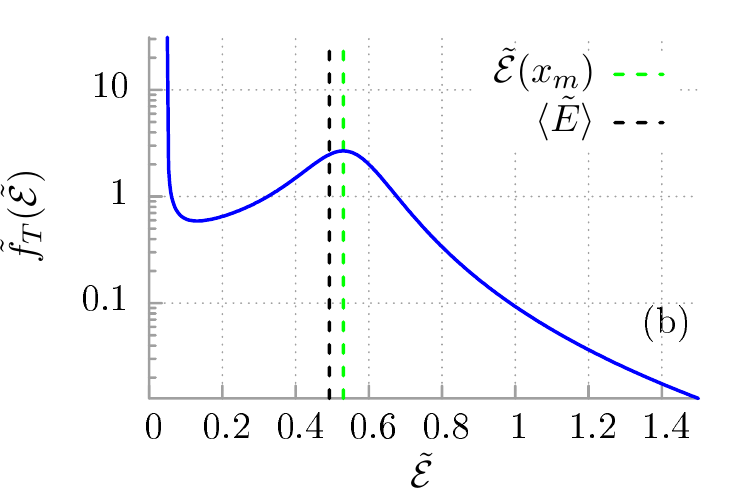}
	\caption{(color online): Drude model. Panel a: The rescaled frequency probability distribution $\tilde{\mathbb{P}}_D(x) = (1/\tau_c)\mathbb{P}_D(x/\tau_c)$ with $x = \omega \tau_c$ is presented for the fixed value of the dimensionless parameter $\alpha = \tau_v/\tau_c = 0.1$.
	 Panel b: The corresponding energy probability distribution $\tilde{f}_{\tilde{T}}(\tilde{\mathcal{E}}) = (\hbar/\tau_c) f_T(\hbar\tilde{\mathcal{E}}/\tau_c)$ is depicted with $\tilde{\mathcal{E}} = \tau_c \mathcal{E}/\hbar$ for the same value of $\alpha$ and temperature $\tilde{T} = \tau_c k_B T/\hbar = 0.1$.}
	\label{fig2}
\end{figure}

In Fig. \ref{fig2} the frequency probability distribution $\mathbb{P}_D(\omega)$ and the energy probability distribution $f_T(\mathcal{E})$ for the Drude model is depicted in the regime where the probability density $\mathbb{P}_D(\omega)$ exhibits a maximum at some value $\omega = \omega_m$, i.e. $\mathbb{P}_D'(\omega_m)=0$ (the prime denotes differentiation with respect to the argument of the function). For the Drude model its value can be analytically evaluated from Eq. (\ref{P_D}) and the result reads   
\begin{equation} \label{max}
\omega_m = \omega_0 \sqrt{1-\alpha}, \quad \omega_0 = \sqrt{\frac{\varepsilon \mu_0}{2}}, 
\quad \alpha < 1. 
\end{equation}
Hence, the distribution $\mathbb{P}_D(\omega)$ displays the non-monotonic character only when $\alpha < 1$. It is the case when the memory time $\tau_c$ is long enough or/and the particle-thermostat coupling constant $\gamma_0$ is sufficiently strong. In other words, the dynamics is pronouncedly non-Markovian and the thermodynamic equilibrium state is far from the Gibbs canonical one. When $\tau_c$ or/and $\gamma_0$ decreases the maximum of $\mathbb{P}_D(\omega)$ disappears. Let us note that the maximum of the corresponding energy probability distribution $f_T(\mathcal{E})$ is not for the value \mbox{$\mathcal{E}(\omega_m) = (\hbar \omega_m/4) \coth\left(\hbar \omega_m/ 2k_BT\right)$} but for the different value $\mathcal{E}_M \equiv \mathcal{E}(\omega_M)$ determined by the condition $f'_T(\mathcal{E})=0$ which leads to the relation 
\begin{equation} \label{Emax}
\mathbb{P}'(\omega_M) \mathcal{E}'(\omega_M) -  \mathbb{P}(\omega_M)\mathcal{E}''(\omega_M) =0, 
\end{equation}
where $\omega=\omega(\mathcal{E})$ is the inverse of the function given by Eq. (\ref{ho}), c.f. Fig. 1. However, because $\mathcal{E}(\omega) \approx (\hbar/4) \omega$ for large values of $\omega$, therefore $\mathcal{E}''(\omega) \approx 0$ and the condition (\ref{Emax}) can be approximated by $\mathbb{P}'(\omega_M) = 0$. Then the exact maximum $\mathcal{E}_M$ of $f_T(\mathcal{E})$ is at the value very close to $\mathcal{E}(\omega_m)$. It is the case presented in panel (b) of Fig. 2 where these two values cannot be  graphically discriminated.
\begin{figure}[t]
	\centering
	\includegraphics[width=0.49\linewidth]{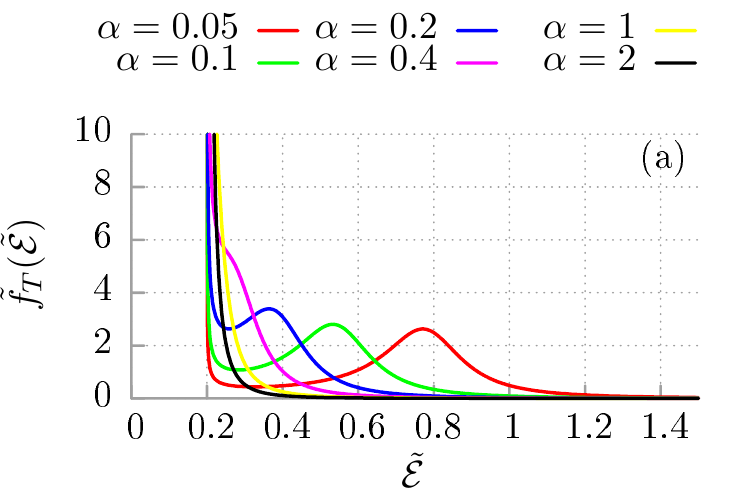}
	\includegraphics[width=0.49\linewidth]{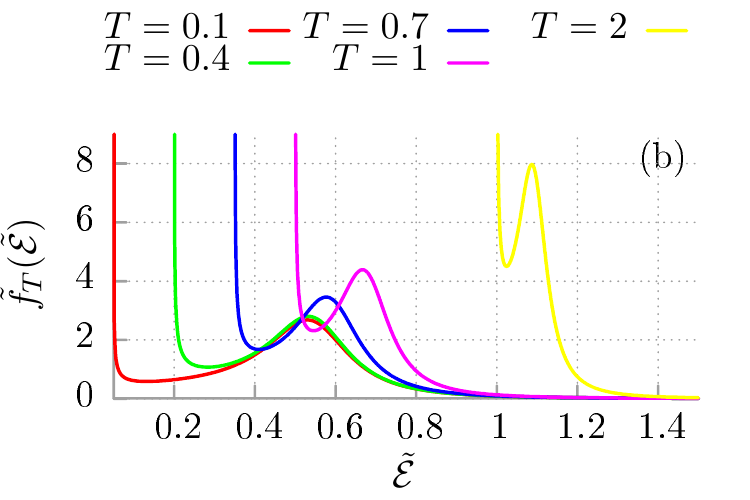}\\
	\includegraphics[width=0.49\linewidth]{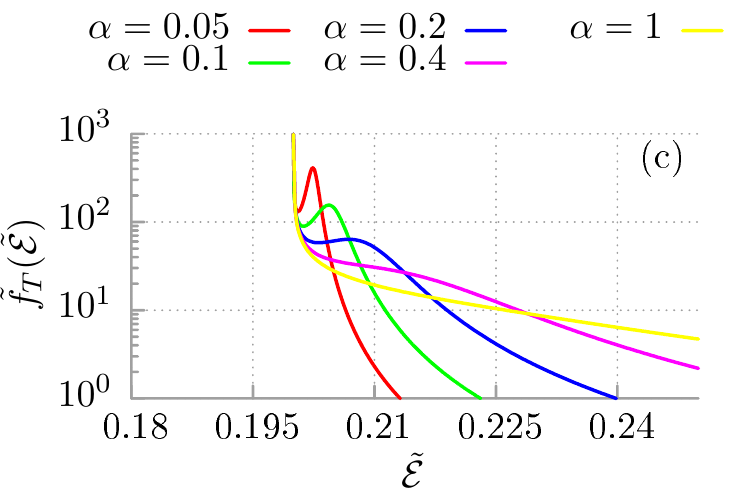}
	\caption{(color online) Drude model. Panel a: The rescaled energy probability distribution $\tilde{f}_{\tilde{T}}(\tilde{\mathcal{E}})$ is depicted for selected values of $\alpha=\tau_v/\tau_c=M/\tau_c \gamma_0$ for $\tilde{T} = 0.4$. The memory time $\tau_c$ is fixed and $\tau_v=M/\gamma_0$ is changed. Panel b: Impact of temperature for fixed $\alpha = 0.1$.
Panel c: The influence of the memory time $\tau_c$ on the same characteristics $\tilde{f}_{\tilde{T}}(\tilde{\mathcal{E}})$ for the fixed $\tau_v$ and temperature $\tilde{T} = 0.4$.}
	\label{fig3}
\end{figure}

In panel (a) of Fig. 3 we depict the impact of the dimensionless parameter $\alpha = M/\tau_c \gamma_0$ on the energy probability distribution $f_T(\mathcal{E})$ at fixed dimensionless temperature $T = 0.4$. Under such a scaling procedure the memory time $\tau_c$ is fixed and $\tau_v=M/\gamma_0$ is changed. It follows that for small values of the parameter $\alpha$, or equivalently for strong coupling $\gamma_0$, the energy probability distribution is notably peaked in the region of larger energies. When $\alpha$ increases (the coupling constant decreases) the most probable value of the thermostat oscillator energy $\mathcal{E}$ is shifted towards smaller values and tends to $\mathcal{E}_0$. For very weak coupling $f_T(\mathcal{E})$ is a monotonically decreasing function of $\mathcal{E}$. While in the present case we have varied the parameter $\alpha$ with fixed temperature $T$, as the next step of our analysis we keep $\alpha = 0.1$ and study the impact of $T$ on the energy probability distribution $f_T(\mathcal{E})$. The corresponding results are presented in Fig. 3 (b). When $T$ increases the support of $f_T(\mathcal{E})$ and the maximum $\mathcal{E}_M$ of $f_T(\mathcal{E})$ are both shifted towards greater energies. Moreover, $f_T(\mathcal{E}_M)$ also grows while at the same time $\mathcal{E}_M$ approaches the minimal value $\mathcal{E}_0$ and finally at high temperature $f_T(\mathcal{E})$ is monotonically decreasing function of $\mathcal{E}$. In panel (c) of Fig. 3 we show the influence of the memory time $\tau_c$ via the parameter $\alpha$ with fixed $\tau_v$ and $T = 0.4$. It means that if $\alpha$ increases $\tau_c$ decreases. For long memory time (small $\alpha$) the dynamics is non-Markovian and the probability density $f_T(\mathcal{E})$ exhibits a local maximum. On the other hand, for vanishing correlation time $\tau_c \to 0$ the distribution $f_T(\mathcal{E})$ becomes a monotonically decreasing function of $\mathcal{E}$.

Finally, in Fig. 4 we compare two dissipation mechanisms, namely the Drude model and the Lorentzian decay of the memory kernel. One can observe that for the Drude mechanism the most probable energy of the thermostat oscillators is greater than for the Lorentzian dissipation. The role of other parameters of the system is similar for both mechanisms, however, the maximum of $f_T(\mathcal{E})$ for the Lorentzian memory function is always at smaller values of $\mathcal{E}$ than for the Drude function.
\begin{figure}[t]
	\centering
	\includegraphics[width=0.49\linewidth]{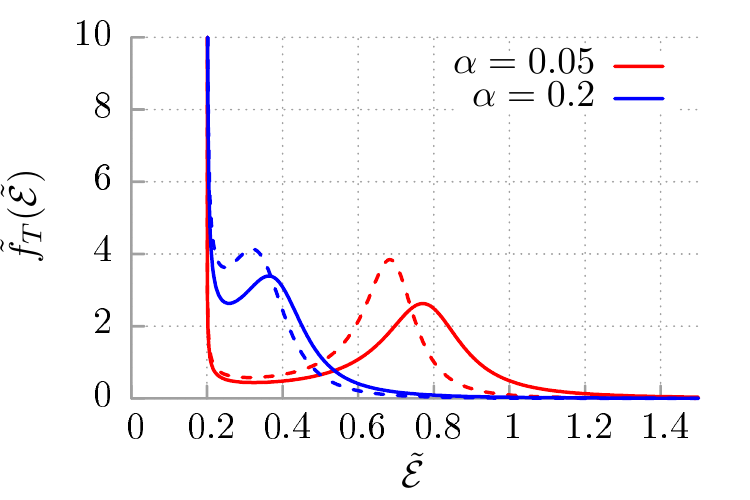}
	\caption{(color online): Two mechanisms of dissipation: Drude (solid lines) and Lorentzian (dashed lines) ones. Two corresponding energy probability distributions are compared for selected  values of the dimensionless parameter $\alpha$ ($M$ or $\gamma_0$ can be changed and $\tau_c$ is fixed) and fixed rescaled temperature $\tilde{T}=0.4$.}
	\label{fig4}
\end{figure}

We conclude that energy probability distribution $f_T(\mathcal{E})$ carries information about quantumness of the environment of the analysed system. In particular, in the classical limit of high temperature it tends to deterministic distribution with the definite energy $\mathcal{E}_0$. It means that all harmonic oscillators building thermostat have the same energy $\mathcal{E}_0$ on average which is exactly equal to the mean energy of the Brownian particle. This is what we expect from the energy equipartition theorem. In contrast, for lower temperatures quantumness of thermostat is visible as a decaying energy probability distribution $f_T(\mathcal{E})$. Moreover, in the limit of strong system-thermostat coupling and/or non-Markovian dynamics the density $f_T(\mathcal{E})$ may be even a non-monotonic function with the local maximum which is dependent on the parameters of the dissipation function $\gamma(t)$ and temperature $T$.
\begin{figure}[t]
	\centering
	\includegraphics[width=0.49\linewidth]{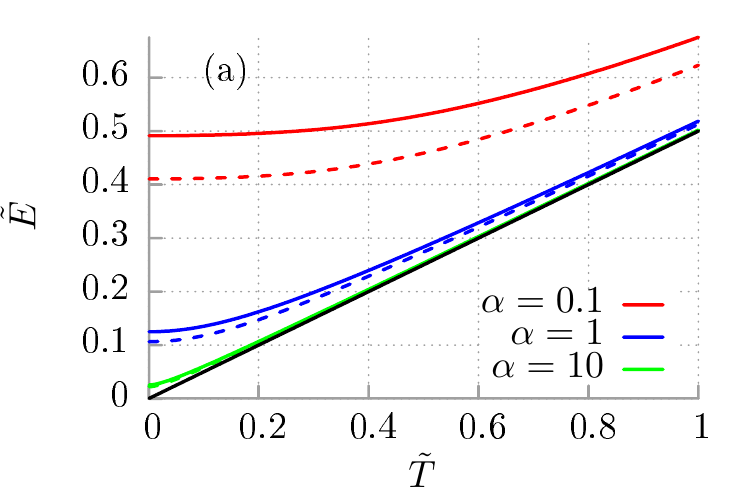}
	\includegraphics[width=0.49\linewidth]{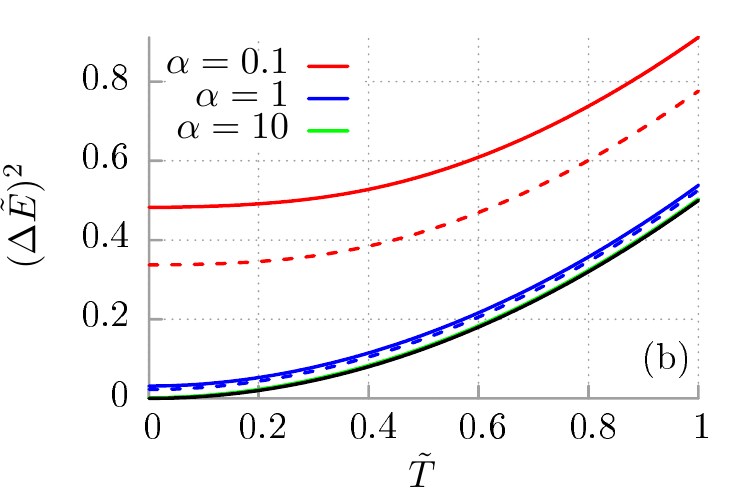}
	\caption{(color online): Panel a: Average energy $\langle \tilde{E} \rangle$ of the Brownian particle versus temperature $\tilde{T}$ of thermostat for selected values of the parameter $\alpha$. Panel b: Fluctuations of energy $(\Delta \tilde{E})^2$ of the Brownian particle depicted as a function of temperature $\tilde{T}$ for different $\alpha$. Solid and dashed lines are for the Drude and Lorentzian dissipation mechanism, respectively. The black line shows the corresponding classical result. In the presented scaling variation of the parameter $\alpha$ means a change of the characteristic time $\tau_v$ with the fixed $\tau_c$.}
	\label{fig5}
\end{figure}
\section{Average energy and its fluctuations}
Let us now discuss the second statistical moment of the Brownian particle energy, namely, 
\begin{equation}
	\langle E^2 \rangle = \int_0^\infty E^2 \,f(E)\,dE.
\end{equation}
To calculate this quantity it is important to note that in the stationary state the Brownian particle momentum is expressed as a convolution of the response function $R(t)$ and the random force $\eta(t)$
\begin{equation}
\lim_{t \to \infty} p(t) = \lim_{t \to \infty} \int_0^t R(t-u)\eta(u)du.
\end{equation}
Statistical characteristics of the random perturbation $\eta(t)$ are analogous to a classical stationary Gaussian stochastic process which models thermal equilibrium noise. Hence $\eta(t)$ is a Gaussian operator representing quantum counterpart of thermal noise. Consequently, the statistical characteristics of the particle momentum is also Gaussian implying that
\begin{equation}
\langle p^4 \rangle = 3\langle p^2 \rangle^2
\end{equation}
which immediately translates to
\begin{equation}
\langle E^2 \rangle = 3\langle E \rangle^2.
\end{equation}
From these relations one can analyse mean energy $\langle E \rangle$ and its fluctuations $(\Delta E)^2 = \langle E^2 \rangle - \langle E \rangle ^2 = 2 \langle E \rangle^2$. In the classical case, from Eq. (\ref{Ecl}) it follows that 
\begin{eqnarray} \label{mcl}
\langle E \rangle_{cl} = \frac{1}{2} k_B T, \quad   
\Delta E_{cl} = \frac{1}{\sqrt{2}} k_B T.
\end{eqnarray}
The mean value is in accordance with the energy equipartition theorem. Note that fluctuations $\Delta E_{cl}$ quantified by the root mean square deviation of the energy from its average are greater than $\langle E \rangle_{cl}$. Moreover, both $\langle E \rangle_{cl}$ and $\Delta E_{cl}$ are linearly increasing functions of temperature $T$. 

In the quantum case the mean energy has been analysed in our previous papers \cite{bialasPRA,bialasJPA,bialasSCIREP}. Here, in Fig. 5 we additionally present fluctuations of the energy of the Brownian particle.  In particular, we observe that both the average energy $\langle E \rangle$ and its fluctuations $(\Delta E)^2$ are monotonically increasing function of temperature. Moreover, if the parameters $\alpha$ grows, meaning that the coupling between the particle and thermostat becomes weaker, again either $\langle E \rangle$ or $(\Delta E)^2$ progressively converge to the corresponding result which is indicated by the black solid line. On the other hand, the strong coupling (smaller $\alpha$) causes an increase of the mean energy and its fluctuations. In the regime of moderate to strong particle-thermostat interaction clear difference between two dissipation mechanisms are observed. For the Lorentzian memory kernel the mean energy as well as fluctuations are smaller than for the Drude model. Finally, we now mention the impact of the correlation time $\tau_c$ on these characteristics (not depicted). The difference between the Drude and the Lorentzian dissipation mechanism becomes negligible in the limit of large $\tau_c$. For smaller values of the correlation time the mean energy as well as its fluctuations are greater for the Drude model. Overall, the latter two quantities grows as $\tau_c$ becomes shorter.
\section{Specific heat}
Finally, in this section we want to analyse the specific heat of a free quantum Brownian particle. For a quantum system weakly coupled to thermostat the heat capacity $C$ can be obtained from the relation
\begin{eqnarray} \label{capa}
C = \frac{\partial U}{\partial T}
\end{eqnarray}
where $U$ is the internal energy
\begin{eqnarray} \label{inter}
U = -\frac{\partial }{\partial \beta} \ln (Z)
\end{eqnarray}
and $Z$ is the partition function of the system. Unfortunately, there is no consensus on the notion of the partition function for quantum systems that are strongly coupled to thermostat. In literature one can find the following definition: the free energy of the system of interest is the free energy of the total system (system + thermostat + interaction) minus the free energy of thermostat in the absence of the system. As a result the partition function of such a system is a ratio of the partition function of the total system and thermostat alone. However, it has been demonstrated \cite{PT} that this definition yields negative value of the specific heat and density of states of the system. For this reason, instead we use the thermodynamic notion of specific heat which is based on the mean energy of the particle, namely
\begin{equation}
C = \frac{\partial\langle E \rangle}{\partial T}
\end{equation}
Exploiting Eqs. (\ref{Ek}) and (\ref{ho}) we arrive at the formula
\begin{equation}\label{C}
C = \frac{1}{2} \int_0^{\infty} c_0(\omega) \mathbb{P}(\omega)  d\omega 
\end{equation}
where $c_0(\omega)$ is the specific heat of a quantum oscillator in a Gibbs canonical state (see Eq. (49.4) in the Landau-Lifshitz book \cite{landau})
\begin{equation}\label{c0}
c_0 =  k_B \left(\frac{\hbar \omega}{k_B T} \right)^2
 \frac{e^{\hbar \omega/k_B T}}{\left(e^{\hbar \omega/k_B T} - 1 \right)^2}. 
\end{equation}
The prefactor $1/2$ is because of one degree of freedom for the Brownian particle while for the oscillator it is two. The form expressed by the frequency distribution $\mathbb{P}(\omega)$ is more convenient than one using the energy probability density $f_T(\mathcal{E})$ which is not known in the analytical form.
%
\begin{figure}[t]
	\centering
	\includegraphics[width=0.49\linewidth]{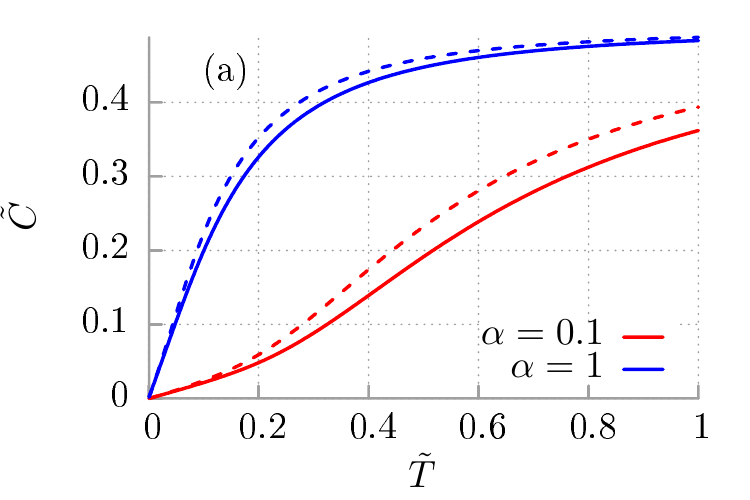}
	\includegraphics[width=0.49\linewidth]{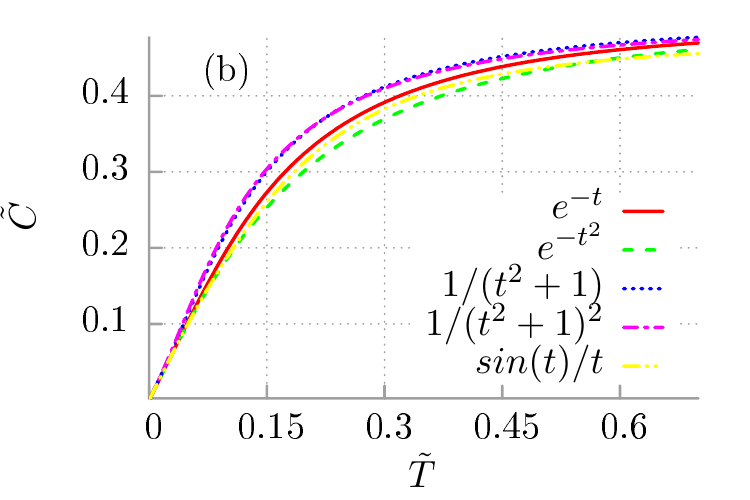}
	\caption{(color online): Panel a: Specific heat $\tilde{C} = C/k_B$ as a function of temperature $\tilde{T} = \tau_c k_B T/\hbar$ is shown for the Drude as well as the Lorentzian dissipation mechanism and two values of the parameter $\alpha$ with the fixed $\tau_c$. Panel b: Comparison of the dependence of the specific heat $\tilde{C}$ on temperature $\tilde{T}$ is depicted for several models of the memory kernel $\gamma(t)$ with $\alpha = 1$.}
	\label{fig6}
\end{figure}

In panel (a) of Fig. 6 we present the specific heat of a free quantum Brownian particle $C$ as a function of temperature for the Drude as well as the Lorentzian model of the dissipation mechanism and different values of the parameter $\alpha$. The scaling is such that the memory time $\tau_c$ is fixed meaning that small $\alpha$ corresponds to the strong coupling limit whereas large $\alpha$ yields the opposite scenario. The observation is that the heat capacity $C$ grows as $\alpha$ is decreasing. Curiously, apart from the very low temperature regime the dissipation mechanism noticeably affects $C$. The latter quantity is greater for the Lorentzian model than the Drude one. In panel (b) of the same figure we take a closer look at the impact of the memory function on the heat capacity of the Brownian particle. In particular, for the fixed $\alpha = 1$ we depict temperature dependence of this quantity for various classes of dissipation mechanisms ranging from the exponential to algebraic and finally the Debye type model, see the Appendix. The general conclusion is that the memory kernel influences the heat capacity especially for moderate to higher temperature regimes. The smallest values are obtained for the Debye model whereas the largest are for both the algebraic and Lorentzian ones since in the case of these two differences are barely visible. Finally, we comment on the role of the memory time $\tau_c$ on the heat capacity (not depicted). Regardless of the model of dissipation this quantity grows for longer correlation time $\tau_c$.
\section{Summary}
In this paper we have returned to the well known problem of quantum Brownian particle. We formulated it in terms of the generalized quantum Langevin equation for a free particle interacting with an infinite number of independent oscillators forming thermal reservoir. It allowed us to analyse average energy of the Brownian particle and its fluctuations as well as the specific heat of the system. In particular, we represented the mean energy of the particle in the superstatistical way as an averaged kinetic energy per one degree of freedom of the thermostat oscillators. The averaging is twofold: (i) over the thermal equilibrium Gibbs state for the thermostat oscillators and (ii) over energies of those thermostat oscillators. The latter is according to the energy probability distribution which carries information about quantumness of the environment of the studied system. We analysed its dependence on various dissipation mechanisms expressed as different forms of the memory kernel in the generalized Langevin equation. Moreover, we studied impact of the system-thermostat coupling strength, the memory time and temperature on the energy probability distribution. Then, we turned to the analysis of the fluctuations of the energy of the quantum Brownian particle and revealed influence of the above parameters on this quantity. Last but not least, the superstatistical representation of the mean energy of the particle allowed us to easily study the specific heat of the system for a whole range of different dissipation kernels. We uncovered the similarities as well as discrepancies between them and discussed impact of the system-thermostat coupling strength and the memory time on the specific heat of the Brownian particle. 

The quantum law for energy partition in the present formulation turned out to be conceptually simple yet very powerful tool for analysis of quantum open systems. We hope that our work will stimulate its further successful applications.
\section*{Acknowledgement}
J. S. was supported by the Grant NCN 2017/26/D/ST2/00543. J. {\L}. was supported by the Grant NCN 2015/19/B/ST2/02856.
\appendix
\section{Solution of the Langevin equation (\ref{GLE2})}
Eq. (\ref{GLE2}) is a linear integro-differential equation for the momentum operator $p(t)$.  
Because its integral part is a convolution, it can be solved by the Laplace transform method yielding 
\begin{equation}\label{pL} 
z \hat p_L(z) -p(0) + \frac{1}{M} \hat \gamma_L(z) \hat p_L(z) = - \hat \gamma_L(z) x(0) + \hat \eta_L(z),   
\end{equation}
where $\hat p_L(z)$,  $\hat \gamma_L(z)$ and $ \hat \eta_L(z)$ are the Laplace transforms of $p(t), \gamma(t)$ and $\eta(t)$, respectively (see  Eq. (\ref{fL}). The operators $p(0)$ and $x(0)$ are the momentum and coordinate operators of the Brownian particle at time $t=0$. 
From this equation it follows that  
\begin{equation}\label{pLL} 
\hat p_L(z) = \hat{R}_L(z) p(0) - \hat{R}_L(z)\hat \gamma_L(z) x(0) + \hat{R}_L(z)\hat \eta_L(z), 
\end{equation}
where
\begin{equation}\label{RL2} 
\hat{R}_L(z) = \frac{M}{Mz + \hat \gamma_L(z)}. 
\end{equation}  
 The inverse Laplace transform of (\ref{pLL}) gives the solution $p(t)$ for the momentum of the Brownian particle, namely, 
\begin{eqnarray}\label{p2(t)} 
p(t) = R(t)p(0) - \int_0^t du\; R(t-u) \gamma(u)x(0) +  \int_0^t du\; R(t-u) \eta(u),   
\end{eqnarray}
where the response function $R(t)$ is the inverse Laplace transform of the function $\hat{R}_L(z)$ in Eq. (\ref{RL2}).  
Because statistical properties of thermal noise $\eta(t)$ are specified, all statistical characteristics of the particle momentum $p(t)$ can be calculated, in particular its kinetic energy.  
\section{Dissipation mechanisms}
In this paper, we consider the following forms of the dissipation function: the Drude
\begin{equation}
	\label{g-drudea}
	\gamma_D(t) = \frac{\gamma_0}{2 \tau_c}e^{-t/\tau_c},
\end{equation}
the Lorentzian
\begin{equation}
	\label{g-lorentz}
\gamma_L(t) = \frac{\gamma_0}{\pi}\, \frac{\tau_c}{t^2 + \tau_c^2},
\end{equation}
the Gaussian
\begin{equation}
	\label{g-gaussian}
\gamma_G(t) = \frac{\gamma_0}{\sqrt{\pi} \tau_c}\, \mbox{e}^{-t^2/\tau_c^2},
\end{equation}
the algebraic
\begin{equation}
	\label{g-algebraic-n}
	\gamma_{A}(t) = \frac{1}{2}\frac{\gamma_0 \tau_c}{(t + \tau_c)^2},
\end{equation}
and finally the Debye model
\begin{equation}
	\label{g-sin}
	\gamma_S(t) = \frac{\gamma_0}{\pi}\frac{\sin{(t/\tau_c)}}{t},  
\end{equation}
where $\gamma_0$ is the particle-thermostat coupling strength and  $\tau_c$ is the memory time.


\section*{References}
\begin{thebibliography}{99}
\bibitem{maga} Magalinskij V B 1959 J. Exp. Theor. Phys. 36 1942

\bibitem{uler} Ullersma P 1966 Physica 32 27

\bibitem{caldeira} Caldeira A O Leggett A J 1983 Ann. Phys. (N.Y.) 149 374

\bibitem{ford} Ford G W and Kac M 1987 J. Stat. Phys. 46 803

\bibitem{gaussian} De Smedt P, D\"urr D and Lebowitz J L 1988 Commun. Math. Phys. 120 195

\bibitem{et2} Ford G W, Lewis J T and O'Connell R F 1988 Phys. Rev. A 37 4419




\bibitem{weis} Weiss U 2008 \textit{Quantum Dissipative Systems} (World Scientific: Singapore)

\bibitem{bialasPRA} Spiechowicz J, Bialas P and {\L}uczka J 2018 Phys. Rev. A 98 052107

\bibitem{bialasJPA} Bialas P, Spiechowicz J and {\L}uczka J 2019 J. Phys. A: Math. Theor. 52 15LT01

\bibitem{feynman} Feynman R P 1972 \textit{Statistical Mechanics} (Westview Press, USA, PA) 

\bibitem{call} Callen H B and Welton T A 1951 Phys. Rev. 83 34

\bibitem{zubarev} Zubarev D N 1974 \textit{Nonequilibrium statistical thermodynamics} (New York, Consultants Bureau)

\bibitem{landau} Landau L D and Lifshitz E M 1980 \textit{Statistical Physics, Part 1} (Butterworth-Heinemann, 3rd ed.)

\bibitem{beck} Beck C and Cohen E 2003 Physica A 322 267

\bibitem{metzler} Chechkin A, Seno F, Metzler R and Sokolov I M 2017 Phys. Rev. X 7 021002

\bibitem{magdziarz} Slezak J, Metzler R and Magdziarz M 2018 New J. Phys. 20 023026

\bibitem{bialasSCIREP} Bialas P, Spiechowicz J and {\L}uczka J 2018 Sci. Rep. 8 16080

\bibitem{PT} Hanggi P, Ingold G L and Talkner P 2008 New J. Phys. 10 115008



















\end{thebibliography}
\end{document}